# HYPERGRAPH PARTITIONING THROUGH VERTEX SEPARATORS ON GRAPHS

ENVER KAYAASLAN*, ALI PINAR†, ÜMIT ÇATALYÜREK‡, AND CEVDET AYKANAT§

**Abstract.** The modeling flexibility provided by hypergraphs has drawn a lot of interest from the combinatorial scientific community, leading to novel models and algorithms, their applications, and development of associated tools. Hypergraphs are now a standard tool in combinatorial scientific computing. The modeling flexibility of hypergraphs however, comes at a cost: algorithms on hypergraphs are inherently more complicated than those on graphs, which sometimes translate to nontrivial increases in processing times. Neither the modeling flexibility of hypergraphs, nor the runtime efficiency of graph algorithms can be overlooked. Therefore, the new research thrust should be how to cleverly trade-off between the two. This work addresses one method for this trade-off by solving the hypergraph partitioning problem by finding vertex separators on graphs. Specifically, we investigate how to solve the hypergraph partitioning problem by seeking a vertex separator on its net intersection graph **(NIG)**, where each net of the hypergraph is represented by a vertex, and two vertices share an edge if their nets have a common vertex. We propose a vertex-weighting scheme to attain good node-balanced hypergraphs, since NIG model cannot preserve node balancing information. Vertex-removal and vertex-splitting techniques are described to optimize cutnet and connectivity metrics, respectively, under the recursive bipartitioning paradigm. We also developed an implementation for our GPVS-based HP formulations by adopting and modifying a state-of-the-art GPVS tool onmetis. Experiments conducted on a large collection of sparse matrices confirmed the validity of our proposed techniques.

**AMS subject classifications.**

**1. Introduction.** A hypergraph is a generalization of a graph, since it replaces edges that connect only two vertices, with hyperedges (nets) that can connect multiple vertices. This generalization provides a critical modeling flexibility that allows accurate formulation of many important problems in combinatorial scientific computing. Our initial motivations for hypergraph models were accurate modeling of the nonzero structure of unsymmetric and rectangular sparse matrices to minimize communication volume for iterative solvers [8, 9, 11, 12, 13, 23, 5, 55, 56, 57, 58, 59] and permutation to block-angular form for coarse-grain parallelism [3]. The real impact of these works turned out to be the introduction of hypergraph models to the combinatorial scientific computing community. Since then, the modeling power of hypergraphs appealed to many researchers and was applied to a wide variety of parallel and distributed computing applications such as data aggregation [15], image-space parallel direct volume rendering [7], parallel mixed integer linear programming [53], data declustering for multi-disk databases [38, 42], scheduling file-sharing tasks in heterogeneous master-slave computing environments [33, 34, 37], and work-stealing scheduling [60]. Hy-pergraphs were also applied to applications outside the parallel computing domain such as road network clustering for efficient query processing [19, 20, 21], pattern-based data clustering [43], reducing software development and maintenance costs [4], processing spatial join operations [51], and improving locality in memory or cache performance [1, 52, 61]. Hypergraphs and hypergraph partitioning are now standard tools of combinatorial scientific computing.

---

*Computer Engineering Department, Bilkent University, Ankara, Turkey (enver@cs.bilkent.edu.tr),

†Sandia National Laboratories, Livermore, CA (apinar@sandia.gov). The work of this author is partially supported by DOE Office of Advanced Scientific Computing Research, Applied Mathematics program.

‡Departments of Biomedical Informatics and Electrical & Computer Engineering, The Ohio State University (umit@bmi.osu.edu). This work of this author is partially supported by the U.S. DOE SciDAC Institute Grant DE-FC02-06ER2775 and by the U.S. National Science Foundation under Grants CNS-0643969, OCI-0904809, and OCI-0904802.

§Computer Engineering Department, Bilkent University, Ankara, Turkey (aykanat@cs.bilkent.edu.tr).

Increasing popularity of hypergraphs has been going hand in hand with the development of effective hypergraph partitioning (HP) tools: wide applicability of hypergraphs motivated development of fast HP tools, and availability of effective HP tools motivated further applications. This virtuous cycle produced sequential HP tools such as hMeTiS [32], PaToH [10] and Mondriaan [59], and parallel HP tools such as Parkway [54] and Zoltan [22], all of which adopt the multilevel framework successfully. While these tools provide good performances both in terms of solution quality and processing times, they are hindered by the inherent complexity of dealing with hypergraphs. Algorithms on hypergraphs are more difficult both in terms of computational complexity and runtime performance, since operations on nets are performed on sets of vertices as opposed to pairs of vertices as in graphs. The wide interest over the last decade has proven the modeling flexibility of hypergraphs to be essential, but the runtime efficiency of graph algorithms cannot be overlooked, either. Therefore, we believe that the new research thrust should be how to cleverly trade-off between the modeling flexibility of hypergraphs and the practicality of graphs.

How can we solve problems that are most accurately modeled with hypergraphs using graph algorithms without sacrificing too much from what is really important for the application? This question has been asked before, and the motivation was either theoretical [29] or practical [14, 28] when the absence of HP tools behest these attempts. This earlier body of work investigated the relation between HP and graph partitioning by edge separator (GPES), and achieved little success. Today, we are facing a more difficult task, as effectiveness of available HP tools sets high standards for novel approaches. On the other hand, we can draw upon the progress on related problems, in particular the advances in tools for graph partitioning by vertex separator (GPVS), which is the main theme of this work.

We investigate solving the HP problem by finding vertex separator on the *net intersection graph* (NIG) of the hypergraph. In the NIGof a hypergraph, each net is represented by a vertex and and each vertex of the hypergraph is replaced with a clique of the nets connecting that vertex. A vertex separator on this graph defines a net separator for the hypergraph. This model has been initially studied for circuit partitioning [2]. While faster algorithms can be designed to find vertex separators on graphs, the NIG model has the drawback of attaining balanced partitions. Once vertices of the hypergraphs are replaced with cliques, it will be impossible to preserve the vertex weight information accurately. Therefore, we can view the NIG model as a way to trade off computational efficiency with exact modeling power.

What motivates us to investigate NIGs to solve HP problems arising in scientific computing applications is that in many applications, definition of balance cannot be very precise [3, 44, 45] or there are additional constraints that cannot be easily incorporated into partitioning algorithms and tools [47]; or partitioning is used as part of a divide-and-conquer algorithm [46]. For instance, hypergraph models can be used to permute a linear program (LP) constraint matrix to a block angular form for parallel solution with decomposition methods. Load balance can be achieved by balancing subproblems during partitioning. However, it is not possible to accurately predict solution time of an LP, and equal sized subproblems only increase the likelihood of computational balance. Hypergraph models have recently been used to find null-space bases that have a sparse inverse [46]. This application requires finding a column-space basis $B$ as a submatrix of a sparse matrix $A$, so that $B^{-1}$ is sparse. Choosing $B$ to have a block angular form limits the fill in $B^{-1}$, but merely a block angular form for $B$ will not be sufficient, since $B$ has to be nonsingular to be a column-space basis for $A$. Enforcing numerical or even structural nonsingularity of subblocks during partitioning is a nontrivial task, if possible, and thus partitioning is used as part of a divide-and-conquer paradigm, where the partitioning phase is followed by a correction phase, if subblocks are non-singular. Both of these cases present examples of applications, where hypergraphs provide effective models, but balance among parts is only weakly defined. As we will show in the experiments, the NIG model can effectively be employed for these applications to achieve high quality solutions in a shorter time. We show that it is easy to enforce a balance criteria on the internal nets of hypergraph partitioning by enforcing vertex balancing during the partitioning of the NIG. However, the NIG model cannot completely preserve the node balancing information in the hypergraph. We propose a weighting scheme for the vertices of NIG, which is quite effective in attaining fairly node balanced partitions of the hypergraph. The proposed vertex balancing scheme for NIG partitioning can be easily enhanced to improve the balancing quality of the hypergraph partitions in a simple post processing phase.

The recursive bipartitioning (RB) paradigm is widely used for multiway HP and known to produce good solution qualities [10, 32]. At each RB step, cutnet removal and cutnet splitting techniques [9] are adopted to optimize the cutsize according to the *cutnet* and *connectivity* metrics, respectively, which are the most commonly used cutsize metrics in scientific and parallel computing [3, 9] as well as VLSI layout design [2, 41]. In this paper, we propose separator-vertex removal and separator-vertex splitting techniques for RB-based partitioning of the NIG, which exactly correspond to the cutnet removal and cutnet splitting techniques, respectively. We also propose an implementation for our GPVS-based HP formulations by adopting and modifying a state-of-the-art GPVS tool used in fill-reducing sparse matrix ordering.

**2. Preliminaries.** In this section, we will provide the basic definitions and techniques that will be adopted in the remainder of this paper.

**2.1. Graph Partitioning.** An undirected graph $\mathcal{G} = (\mathcal{V}, \mathcal{E})$ is defined as a set $\mathcal{V}$ of vertices and a set $\mathcal{E}$ of edges. Every edge $e_{ij} \in \mathcal{E}$ connects a pair of distinct vertices $v_i$ and $v_j$. We use the notation $Adj(v_i)$ to denote the set of vertices adjacent to vertex $v_i$. We extend this operator to include the adjacency set of a vertex subset $\mathcal{V}' \subset \mathcal{V}$, i.e., $Adj(\mathcal{V}') = \{v_j \in \mathcal{V} - \mathcal{V}' : v_j \in Adj(v_i) \text{ for some } v_i \in \mathcal{V}'\}$. Two disjoint vertex subsets $\mathcal{V}_k$ and $\mathcal{V}_\ell$ are said to be adjacent if $Adj(\mathcal{V}_k) \cap \mathcal{V}_\ell \neq \emptyset$ (equivalently $Adj(\mathcal{V}_\ell) \cap \mathcal{V}_k \neq \emptyset$) and non-adjacent otherwise. The degree $d(v_i)$ of a vertex $v_i$ is equal to the number of edges incident to $v_i$, i.e., $d(v_i) = |Adj(v_i)|$. A weight $w(v_i) \geq 0$ is associated with each vertex $v_i$.

An edge subset $\mathcal{E}_S$ is a $K$-way *edge separator* if its removal disconnects the graph into at least $K$ connected components. That is, $\Pi_{ES}(\mathcal{G}) = \{\mathcal{V}_1, \mathcal{V}_2, \ldots, \mathcal{V}_K\}$ is a K-way vertex partition of $\mathcal{G}$ by edge separator $\mathcal{E}_S \subset \mathcal{E}$ if each part $\mathcal{V}_k$ is non-empty; parts are pairwise disjoint; and the union of parts gives $\mathcal{V}$. Edges between the vertices of different parts belong to $\mathcal{E}_S$, and are called *cut (external)* edges and all other edges are called *uncut (internal)* edges.

A vertex subset $\mathcal{V}_S$ is a $K$-way *vertex separator* if the subgraph induced by the vertices in $\mathcal{V} - \mathcal{V}_S$ has at least $K$ connected components. That is, $\Pi_{VS}(\mathcal{G}) = \{\mathcal{V}_1, \mathcal{V}_2, \ldots, \mathcal{V}_K; \mathcal{V}_S\}$ is a K-way vertex partition of $\mathcal{G}$ by vertex separator $\mathcal{V}_S \subset \mathcal{V}$ if each part $\mathcal{V}_k$ is non-empty; all parts and the separator are pairwise disjoint; parts are pairwise non-adjacent; and the union of parts and the separator gives $\mathcal{V}$. The non-adjacency of the parts implies that $Adj(\mathcal{V}_k) \subseteq \mathcal{V}_S$ for each $\mathcal{V}_k$. A vertex $v_i \in \mathcal{V}_k$ is said to be a boundary vertex of part $\mathcal{V}_k$ if it is adjacent to any vertex in $\mathcal{V}_S$. A vertex separator is said to be *narrow* if no subset of it forms a separator, and *wide* otherwise.

The objective of graph partitioning is finding a separator of smallest size subject to a given balance criteria on the weights of the $K$ parts. The weight $W(\mathcal{V}_k)$ of a part $V_k$ is defined as the sum of the weights of the vertices in $\mathcal{V}_k$, i.e.,

$$W(\mathcal{V}_k) = \sum_{v_i \in \mathcal{V}_k} w(v_i) \tag{2.1}$$

and the balance criteria is defined as

$$\max_{1 \leq k \leq K} W(\mathcal{V}_k) \leq (1 + \epsilon) W_{avg} , \quad \text{where} \tag{2.2}$$

$$W_{avg} = \frac{\sum_{k=1}^{K} W(\mathcal{V}_k)}{K}$$

Here, $W_{avg}$ is the weight each part must have in the case of perfect balance, and $\epsilon$ is the maximum imbalance ratio allowed. We proceed with formal definitions for the GPES and GPVS problems, both of which are known to be NP-hard [6].

DEFINITION 2.1 (Problem GPES). *Given a graph $\mathcal{G} = (\mathcal{V}, \mathcal{E})$, an integer $K$, and a maximum allowable imbalance ratio $\epsilon$. The GPES problem is finding a K-way vertex partition $\Pi_{ES}(\mathcal{G}) = \{\mathcal{V}_1, \mathcal{V}_2, \ldots, \mathcal{V}_K\}$ of $\mathcal{G}$ by edge separator $\mathcal{E}_S$ that satisfies the balance criterion given in (2.2) while minimizing the cutsize, which is defined as*

$$cutsize(\Pi_{ES}) = \sum_{e_{ij} \in \mathcal{E}_S} c(e_{ij}), \tag{2.3}$$

*where $c(e_{ij}) \geq 0$ is the cost of edge $e_{ij} = (v_i, v_j)$.*

DEFINITION 2.2 (Problem GPVS). *Given a graph $\mathcal{G} = (\mathcal{V}, \mathcal{E})$, an integer $K$, and a maximum allowable imbalance ratio $\epsilon$. The GPVS problem is finding a K-way vertex partition $\Pi_{VS}(\mathcal{G}) = \{\mathcal{V}_1, \mathcal{V}_2, \ldots, \mathcal{V}_K; \mathcal{V}_S\}$ of $\mathcal{G}$ by vertex separator $\mathcal{V}_S$ that satisfies the balance criterion given in (2.2) while minimizing the cutsize, which is defined as one of*

$$a) \; cutsize(\Pi_{VS}) = \sum_{v_i \in \mathcal{V}_S} c(v_i) \tag{2.4}$$

$$b) \; cutsize(\Pi_{VS}) = \sum_{v_i \in \mathcal{V}_S} c(v_i)(\lambda(v_i) - 1) \tag{2.5}$$

*where $c(v_i) \geq 0$ is the cost of vertex $v_i$.*

In the general GPVS definition given above, both a weight and a cost are associated with each vertex. The weights are used in computing loads of parts for balancing, whereas the costs are utilized in computing the cutsize. In the standard GPVS definitions in the literature, the weights and costs of the vertices are taken as identical. The reason for our general GPVS definition will become clear in Section 3.

In the cutsize definition given in (2.4), each separator vertex incurs its cost to the cutsize, whereas in (2.5), the connectivity of a vertex is considered while incurring its cost to the cutsize. The *connectivity* $\lambda(v_i)$ of a vertex $v_i$ denotes the number of parts connected by $v_i$, where a vertex that is adjacent to any vertex in a part is said to *connect* that part.

The techniques for solving GPES and GPVS problems are closely related. An *indirect* approach to solve the GPVS problem is to first find an edge separator through GPES, and then translate it to any vertex separator. After finding an edge separator, this approach takes vertices adjacent to separator edges as a wide separator to be refined to a narrow separator, with the assumption that a small edge separator is likely to yield a small vertex separator. The wide-to-narrow refinement problem [49] is described as a minimum vertex cover problem on the bipartite graph induced by the cut edges. A minimum vertex cover can be taken as a narrow separator for the whole graph, because each cut edge will be adjacent to any vertex in the vertex cover.

**2.2. Hypergraph Partitioning.** A hypergraph $\mathcal{H} = (\mathcal{U}, \mathcal{N})$ is defined as a set $\mathcal{U}$ of nodes (vertices) and a set $\mathcal{N}$ of nets among those vertices. We refer to the vertices of $\mathcal{H}$ as nodes, to avoid the confusion between graphs and hypergraphs. Every net $n_i \in \mathcal{N}$ connects a subset of nodes, i.e., $n_i \subseteq \mathcal{U}$. The nodes connected by a net $n_i$ are called *pins* of $n_i$ and denoted as $Pins(n_i)$. We extend this operator to include the pin list of a net subset $\mathcal{N}' \subset \mathcal{N}$, i.e., $Pins(\mathcal{N}') = \bigcup_{n_i \in \mathcal{N}'} Pins(n_i)$. The size $s(n_i)$ of a net $n_i$ is equal to the number of its pins, i.e., $s(n_i) = |Pins(n_i)|$. The set of nets that connect a node $u_j$ is denoted as $Nets(u_j)$. We also extend this operator to include the net list of a node subset $\mathcal{U}' \subset \mathcal{U}$, i.e., $Nets(\mathcal{U}') = \bigcup_{u_j \in \mathcal{U}'} Nets(u_j)$. The degree $d(u_j)$ of a node $u_j$ is equal to the number of nets that connect $u_j$, i.e., $d(u_j) = |Nets(u_j)|$. The total number $p$ of pins denote the size of $\mathcal{H}$ where $p = \sum_{n_i \in \mathcal{N}} s(n_i) = \sum_{u_j \in \mathcal{U}} d(u_j)$. A graph is a special hypergraph such that each net has exactly two pins. A weight $w(u_j)$ is associated with each node $u_j$, whereas a cost $c(n_i)$ is associated with each net $n_i$. A weight $w(n_i)$ can also be associated with each net $n_i$ as we will discuss later in this section.

A net subset $\mathcal{N}_S$ is a $K$-way *net separator* if its removal disconnects the hypergraph into at least $K$ connected components. That is, $\Pi_{\mathcal{U}}(\mathcal{H}) = \{\mathcal{U}_1, \mathcal{U}_2, \ldots, \mathcal{U}_K\}$ is a $K$-way node partition of $\mathcal{H}$ by net separator $\mathcal{N}_S \subset \mathcal{N}$ if each part $\mathcal{U}_k$ is non-empty; parts are pairwise disjoint; and the union of parts gives $\mathcal{U}$. In a partition $\Pi_{\mathcal{U}}(\mathcal{H})$, a net that connects any node in a part is said to *connect* that part. The *connectivity* $\lambda(n_i)$ of a net $n_i$ denotes the number of parts connected by $n_i$. Nets connecting multiple parts belong to $\mathcal{N}_S$, and are called *cut (external)* (i.e., $\lambda(n_i) > 1$), and *uncut (internal)* otherwise (i.e., $\lambda(n_i) = 1$). The set of internal nets of a part $\mathcal{U}_k$ is denoted as $\mathcal{N}_k$, for $k = 1, \ldots, K$. So, although $\Pi_{\mathcal{U}}(\mathcal{H})$ is defined as a $K$-way partition on the node set of $\mathcal{H}$, it can also be considered as inducing a $(K+1)$-way partition $\Pi_{\mathcal{N}}(\mathcal{H}) = \{\mathcal{N}_1, \ldots, \mathcal{N}_K; \mathcal{N}_S\}$ on the net set.

As in the GPES and GPVS problems, the objective of HP problem is finding a net separator of smallest size subject to a given balance criteria on the weights of the $K$ parts. The weight $W(\mathcal{U}_k)$ of a part $\mathcal{U}_k$ is defined either as the sum of the weights of nodes in $\mathcal{U}_k$, i.e.,

$$W(\mathcal{U}_k) = \sum_{u_j \in \mathcal{U}_k} w(u_j) \qquad (2.6)$$

or as the sum of weights of internal nets of part $\mathcal{U}_k$, i.e.,

$$W(\mathcal{U}_k) = \sum_{n_i \in \mathcal{N}_k} w(n_i) \tag{2.7}$$

The former and latter part weight computation schemes together with the load balancing criteria given in (2.2) will be referred to here as node and net balancing, respectively. We proceed with formal definition for the HP problem, which is also known to be NP-hard [41].

DEFINITION 2.3 (Problem HP). *Given a hypergraph* $\mathcal{H} = (\mathcal{U}, \mathcal{N})$, *an integer* $K$, *and a maximum allowable imbalance ratio* $\epsilon$. *The HP problem is finding a K-way node partition* $\Pi_{\mathcal{U}}(\mathcal{H}) = \{\mathcal{U}_1, \mathcal{U}_2, \ldots, \mathcal{U}_K\}$ *of* $\mathcal{H}$ *that satisfies the balance criterion given in* (2.2) *while minimizing the cutsize, which is defined as one of*

$$a) \ cutsize(\Pi_{\mathcal{U}}) = \sum_{n_i \in \mathcal{N}_S} c(n_i) \tag{2.8}$$

$$b) \ cutsize(\Pi_{\mathcal{U}}) = \sum_{n_i \in \mathcal{N}_S} c(n_i)(\lambda(n_i) - 1) \tag{2.9}$$

The cutsize metrics given in (2.8) and (2.9) are referred to as the *cut-net* and *connectivity* metrics, respectively, [9, 13, 41].

**3. Formulating the HP Problem as a GPVS Problem.** In this section, we first review the previous work on alternative models for solving the HP problem. Then, we describe our novel and accurate GPVS-based formulations and present the relation between HP and GPVS problems from a matrix theoretical view. Finally, we present our implementation based on adapting a state-of-the-art GPVS tool.

**3.1. Alternative Models for Solving the HP Problem.** As indicated in the survey by Alpert and Kahng [2], hypergraphs are commonly used to represent circuit netlist connections in solving the circuit partitioning and placement problems in VLSI layout design. The circuit partitioning problem is to divide a system specification into clusters to minimize inter-cluster connections. Other circuit representation models were also proposed and used in the VLSI literature including dual hypergraph, clique-net graph (CNG) and net-intersection graph (NIG) [2]. Hypergraphs represent circuits in a natural way so that the circuit partitioning problem is directly described as an HP problem. Thus, these alternative models can be considered as alternative approaches for solving the HP problem.

The dual of a hypergraph $\mathcal{H} = (\mathcal{U}, \mathcal{N})$ is defined as a hypergraph $\mathcal{H}'$, where the nodes and nets of $\mathcal{H}$ become, respectively, the nets and nodes of $\mathcal{H}'$. That is, $\mathcal{H}' = (\mathcal{U}', \mathcal{N}')$ with $Nets(u_i') = Pins(n_i)$ for each $u_i' \in \mathcal{U}'$ and $n_i \in \mathcal{N}$, and $Pins(n_j') = Nets(u_j)$ for each $n_j' \in \mathcal{N}'$ and $u_j \in \mathcal{U}$.

In the CNG model, the vertex set of the target graph is equal to the node set of the given hypergraph. Each net of the given hypergraph is represented by a clique of vertices corresponding to its pins. The multiple edges between two vertices are contracted into a single edge, the cost of which is set equal to the sum of the cost of the edges it represents. If an edge is in the cut set of a GPES then all nets represented by this edge are in the cut set of HP. Ideally, no matter how nodes of a net are partitioned, the contribution of a cut net to the cutsize should always be one in a bipartition when unit net costs are assumed. However, the deficiency of the CNG representation is that it is impossible to achieve such a *perfect* edge-cost assignment of the edges as proved by Ihler et al. [29].

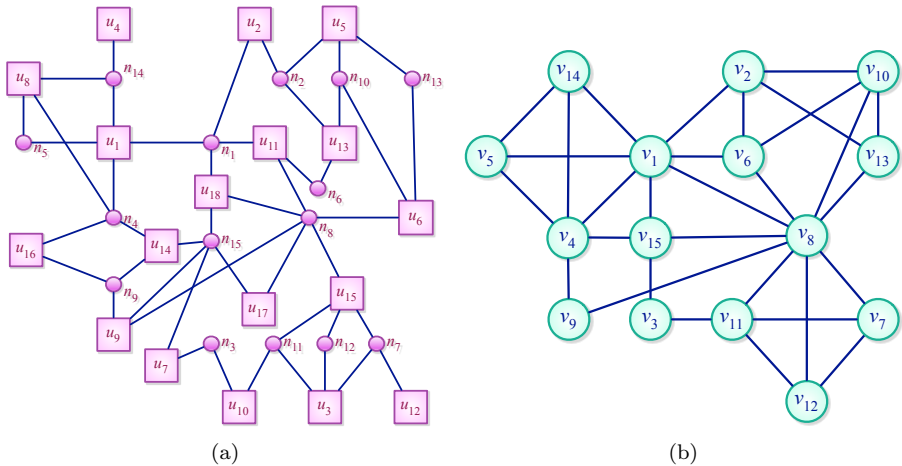

(a)                              (b)

FIG. 3.1. *(a) A sample hypergraph $\mathcal{H}$ and (b) the corresponding NIG representation $\mathcal{G}$.*

In the NIG representation $\mathcal{G} = (\mathcal{V}, \mathcal{E})$ of a given hypergraph $\mathcal{H} = (\mathcal{U}, \mathcal{N})$, each vertex $v_i$ of $\mathcal{G}$ corresponds to net $n_i$ of $\mathcal{H}$. Two vertices $v_i, v_j \in \mathcal{V}$ of $\mathcal{G}$ are adjacent if and only if respective nets $n_i, n_j \in \mathcal{N}$ of $\mathcal{H}$ share at least one pin, i.e., $e_{ij} \in \mathcal{E}$ if and only if $Pins(n_i) \cap Pins(n_j) \neq \emptyset$. So,

$$Adj(v_i) = \{v_j \ : \ n_j \in \mathcal{N} \ \ni \ Pins(n_i) \cap Pins(n_j) \neq \emptyset\}. \qquad (3.1)$$

Note that for a given hypergraph $\mathcal{H}$, NIG $\mathcal{G}$ is well-defined, however there is no unique reverse construction [2]. Figures 3.1(a) and 3.1(b), respectively, display a sample hypergraph $\mathcal{H}$ and the corresponding NIG representation $\mathcal{G}$. In the figure, the sample hypergraph $\mathcal{H}$ contains 18 nodes and 15 nets, whereas the corresponding NIG $\mathcal{G}$ contains 15 vertices and 30 edges.

Both dual hypergraph and NIG models view the HP problem in terms of partitioning nets instead of nodes. Kahng [30] and Cong, Hagen, and Kahng [16] exploited this perspective of the NIG model to formulate the hypergraph bipartitioning problem as a two-stage process. In the first stage, nets of $\mathcal{H}$ are bipartitioned through 2-way GPES of its NIG $\mathcal{G}$. The resulting net bipartition induces a partial node bipartition on $\mathcal{H}$. Because, the nodes (pins) that are connected only by the nets on one side of the bipartition can be unambiguously assigned to that side. However, other nodes may be connected by the nets on both sides of the bipartition. Thus, the second stage involves finding the best completion of the partial node bipartition; i.e., a part assignment for the shared nodes such that the cutsize is minimized. This problem is known as the module (node) contention problem in the VLSI community. Kahng [30] used a winner-loser heuristic [27], whereas Cong et al. [16] used a matching-based (IG-match) algorithm for solving the 2-way module contention problem optimally. Cong, Labio, and Shivakumar [17] extended this approach to $K$-way HP through using the dual hypergraph model. In the first stage, a $K$-way net partition is obtained through partitioning the dual hypergraph. For the second stage, they formulated the $K$-way module contention problem as a min-cost max-flow problem through defining binding factors between nodes and nets, and a preference function between parts and nodes.

Here, we reveal the fact that the module contention problem encountered in the second stage of the NIG-based hypergraph bipartitioning approaches [16, 30] is similar to the wide-to-narrow separator refinement problem encountered in the second stage of the indirect GPVS approaches widely used in nested-dissection based low-fill orderings for sparse matrix factorization. The module contention and separator refinement algorithms effectively work on the bipartite graph induced by the cut edges of a two-way GPES of the NIG representation of hypergraphs and the standard graph representation of sparse matrices, respectively. The winner-loser assignment heuristic [27, 30] used by Kahng [30] is very similar to the minimum-recovery heuristic proposed by Leiserson and Lewis [40] for separator refinement. Similarly, the IG-match algorithm proposed by Cong et al. [16] is similar to the maximum-matching based minimum vertex-cover algorithm [39, 48] used by Pothen, Simon, and Liou [49] for separator refinement. While not explicitly stated in the literature, these net-bipartitioning-based HP algorithms using the NIG model can be viewed as trying to solve the HP problem through an indirect GPVS of the NIG representation.

More recently, Trifunovic and Knottenbelt [54] proposed a coloring-based graph model for partitioning special type of hypergraphs which arise in fine-grain (nonzero-based) partitioning of sparse matrices [13, 11] for parallel matrix vector multiply. In such hypergraphs, each vertex is connected by exactly two nets and their dual hypergraphs are bipartite graphs. A $K$-way edge coloring on this bipartite graph is decoded as a $K$-way partitioning of the nodes (nonzeros) of the original hypergraph. The coloring objective, which is defined in terms of the number of distinct colors incident to the vertices, correctly models the total interprocessor communication volume. Since connectivity cutsize metric (2.9) also correctly models total interprocessor communication volume, the coloring objective exactly models the connectivity cutsize metric (2.9). Although this model is proposed for special type of hypergraphs in which each node is connected by exactly two nets, the model easily extends to more general hypergraphs where nodes are connected by arbitrary number of nets.

**3.2. An Accurate Formulation of HP as GPVS on NIG Model.** We propose a net-partitioning based $K$-way HP algorithm that avoids the module contention problem by describing the HP problem as a GPVS problem through the NIG model. The following theorem lays down the basis for our GPVS-based HP formulation. Let $\mathcal{G} = (\mathcal{V}, \mathcal{E})$ denote the NIG of a given hypergraph $\mathcal{H} = (\mathcal{U}, \mathcal{N})$. The cost of each net $n_i$ of $\mathcal{H}$ is assigned as the cost of the respective vertex $v_i$ of $\mathcal{G}$, i.e., $c(v_i) = c(n_i)$. For brevity of the presentation we assume unit net costs here, but all proposed models and methods generalize to hypergraphs with non-unit net costs.

THEOREM 1. *A $K$-way vertex partition $\Pi_{VS}(\mathcal{G}) = \{\mathcal{V}_1, \dots, \mathcal{V}_K; \mathcal{V}_S\}$ of $\mathcal{G}$ by a narrow vertex separator $\mathcal{V}_S$ induces a $K$-way contention-free net partition $\Pi_{\mathcal{N}}(\mathcal{H}) = \{\mathcal{N}_1 \equiv \mathcal{V}_1, \mathcal{N}_2 \equiv \mathcal{V}_2, \dots, \mathcal{N}_K \equiv \mathcal{V}_K; \mathcal{N}_S \equiv \mathcal{V}_S\}$ of $\mathcal{H}$ by a net separator $\mathcal{N}_S$.*

*Proof.* By definition of GPVS, we have $Adj(\mathcal{V}_k) \cap \mathcal{V}_\ell = \emptyset$ for $1 \leq k < \ell \leq K$. This implies that $Pins(\mathcal{N}_k) \cap Pins(\mathcal{N}_\ell) = \emptyset$ for $1 \leq k < \ell \leq K$. Because, if any two nets $n_i \in \mathcal{N}_k$ and $n_j \in \mathcal{N}_\ell$ shared at least one pin, then there would be an edge $e_{ij}$ between vertices $v_i \in \mathcal{V}_k$ and $v_j \in \mathcal{V}_\ell$ of $\mathcal{G}$, which would correspond to an edge between parts $\mathcal{V}_k$ and $\mathcal{V}_\ell$ of $\Pi_{VS}(\mathcal{G})$ contradicting the definition of GPVS. Therefore, any two nets belonging to two different net parts don't share any pin, thus ensuring the contention-free property of the net partition $\Pi_{\mathcal{N}}(\mathcal{H})$. □

COROLLARY 1. *A $K$-way contention-free net partition of $\mathcal{H}$ by a net separator $\mathcal{N}_S$*

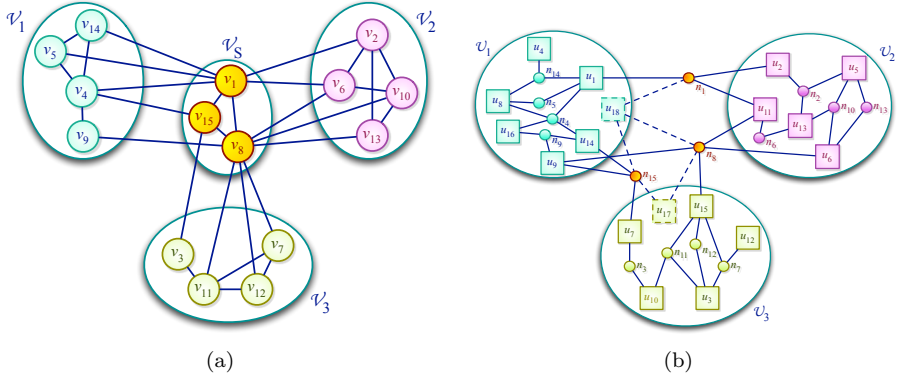

(a)                                (b)

Fig. 3.2. *(a) A 3-way GPVS of the sample NIG given in Figure 3.1(b) and (b)corresponding partitioning of the hypergraph.*

$$\Pi_{\mathcal{N}}(\mathcal{H}) = \{\mathcal{N}_1 \equiv \mathcal{V}_1, \ldots, \mathcal{N}_K \equiv \mathcal{V}_K; \mathcal{N}_S \equiv \mathcal{V}_S\} \tag{3.2}$$

*induces a K-way partial node partition*

$$\Pi'_{\mathcal{U}}(\mathcal{H}) = \{\mathcal{U}'_1 = Pins(\mathcal{N}_1), \ldots, \mathcal{U}'_K = Pins(\mathcal{N}_K)\} \tag{3.3}$$

Let $\mathcal{U}_F$ denote the set of remaining nodes. Note that $\mathcal{U}_F$ also corresponds to the set of nodes that are only connected by the nets of the separator $\mathcal{N}_S$. That is,

$$\mathcal{U}_F = \mathcal{U} - \bigcup_{k=1}^{K} \mathcal{U}'_k = \{u_i \in \mathcal{U} : Nets(u_i) \subseteq \mathcal{N}_S \equiv \mathcal{V}_S\} \tag{3.4}$$

The nodes in $\mathcal{U}_F$ will be referred to here as *free* nodes.

Figure 3.2(a) shows a 3-way GPVS $\Pi_{VS}(\mathcal{G})$ of the sample NIG $\mathcal{G}$ given in Figure 3.1(b). Figure 3.2(b) shows the 3-way partial and complete node partition $\Pi'_{\mathcal{U}}(\mathcal{H})$ of the sample $\mathcal{H}$, which is induced by $\Pi_{VS}(\mathcal{G})$. Partial node partition is displayed with nodes drawn with solid lines, and complete node partition is achieved by adding 2 free nodes (drawn with dashed lines). The sample $\mathcal{H}$ given in Figure 3.1(a) contains only 2 free nodes, which are $u_{17}$ and $u_{18}$. Comparison of Figures 3.2(a) and 3.2(b) illustrates that the separator vertices $v_1, v_8$ and $v_{15}$ of $\Pi_{VS}(\mathcal{G})$ induce the cut nets $n_1, n_8$, and $n_{15}$ of $\Pi'_{\mathcal{U}}(\mathcal{H})$, respectively.

For any arbitrary assignment of free nodes, we can construct a complete node partition in the following form,

$$\Pi_{\mathcal{U}}(\mathcal{H}) = \{\mathcal{U}_1 \supseteq \mathcal{U}'_1, \mathcal{U}_2 \supseteq \mathcal{U}'_2, \ldots, \mathcal{U}_K \supseteq \mathcal{U}'_K\} \tag{3.5}$$

Note that any $K$-way node partition of $\mathcal{H}$ inducing the $(K+1)$-way net partition $\Pi_{\mathcal{N}}(\mathcal{H})$ has to be in the form above.

THEOREM 2. *Given a K-way vertex partition $\Pi_{VS}(\mathcal{G})$ of $\mathcal{G}$ by a narrow vertex separator $\mathcal{V}_S$, any node partition $\Pi_{\mathcal{U}}(\mathcal{H})$ of $\mathcal{H}$ as constructed according to (3.5) induces the $(K+1)$-way net partition $\Pi_{\mathcal{N}}(\mathcal{H}) = \{\mathcal{N}_1 \equiv \mathcal{V}_1, \ldots, \mathcal{N}_K \equiv \mathcal{V}_K; \mathcal{N}_S \equiv \mathcal{V}_S\}$ such* that the connectivity of each cut net in $\mathcal{N}_S$ is greater than or equal to the connectivity of the corresponding separator vertex in $\mathcal{V}_S$.

*Proof.* Let $\Pi_{\mathcal{U}}(\mathcal{H})$ be a node partition of $\mathcal{H}$ as constructed according to (3.5). Consider the net $n_i \in \mathcal{N}_k$ corresponding to a vertex $v_i \in \mathcal{V}_k$ of $\Pi_{VS}(\mathcal{G})$. Since $Pins(n_i) \subseteq \mathcal{U}_k$, $n_i$ will be an internal net of node part $\mathcal{U}_k$ for node partition $\Pi_{\mathcal{U}}(\mathcal{H})$. Consider any net $n_s \in \mathcal{N}_S$ corresponding to separator vertex $v_s \in \mathcal{V}_S$ of $\Pi_{\mathcal{U}}(\mathcal{G})$. Since $\mathcal{V}_S$ is narrow, there exists at least two vertices $v_{j_1} \in \mathcal{V}_{k_1}$ and $v_{j_2} \in \mathcal{V}_{k_2}$ adjacent to $v_s$ such that $k_1 \neq k_2$. Then, corresponding nets $n_{j_1}$ and $n_{j_2}$ are internal nets of $\mathcal{U}_{k_1}$ and $\mathcal{U}_{k_2}$, respectively by Theorem 1. Since the vertices $v_{j_1}$ and $v_{j_2}$ are adjacent to $v_s$ in $\mathcal{G}$, there exists two nodes $u_{h_1}$ and $u_{h_2}$ such that $u_{h_1} \in Pins(n_{j_1}) \cap Pins(n_i)$ and $u_{h_2} \in Pins(n_{j_2}) \cap Pins(n_i)$ by the NIG definition. By the construction of $\Pi_{\mathcal{U}}(\mathcal{H})$, since $Pins(n_{j_1}) \subseteq \mathcal{U}_{k_1}$ and $Pins(n_{j_2}) \subseteq \mathcal{U}_{k_2}$, we have $u_{h_1} \in \mathcal{U}_{k_1}$ and $u_{h_2} \in \mathcal{U}_{k_2}$, which in turn implies that $n_s$ is a cut net for the node partition $\Pi_{\mathcal{U}}(\mathcal{H})$. Therefore $\Pi_{\mathcal{U}}(\mathcal{H})$ induces the net partition $\Pi_{\mathcal{N}}(\mathcal{H}) = \{\mathcal{N}_1 \equiv \mathcal{V}_1, \ldots, \mathcal{N}_K \equiv \mathcal{V}_K; \mathcal{N}_S \equiv \mathcal{V}_S\}$.

Consider the connectivity of the net $n_s \in \mathcal{N}_S$ corresponding to a separator vertex $v_s \in \mathcal{V}_S$ of $\Pi_{VS}(\mathcal{G})$. Since $Pins(\mathcal{N}_k) \subseteq \mathcal{U}_k$, for $1 \leq k \leq K$, if vertex-part $\mathcal{V}_k$ of $\Pi_{VS}(\mathcal{G})$ contains a vertex $v_j \in \mathcal{V}_k$ that is adjacent to $v_s$ then node-part $\mathcal{U}_k$ of $\Pi_{\mathcal{U}}(HP)$ contains a node $u_h$ such that $u_h \in Pins(n_s) \cap Pins(n_j)$. Thus, if a separator vertex $v_s \in \mathcal{V}_S$ connects vertex part $\mathcal{V}_k$, the net $n_s \in \mathcal{N}_S$ also connects node part $\mathcal{U}_k$. The connectivity of net $n_s$ may become strictly greater than that of vertex $v_s$ if $n_s$ connects a free node $u_f$ assigned to a part $\mathcal{U}_\ell$ that is not connected by $n_s$ in partial node partition $\Pi'_{\mathcal{U}}(\mathcal{H})$, i.e., $\mathcal{U}'_\ell \cap Pins(n_s) = \emptyset$. □

COROLLARY 2. *Given a $K$-way vertex partition $\Pi_{VS}(\mathcal{G})$ of $\mathcal{G}$ by a narrow vertex separator $\mathcal{V}_S$, the separator size of $\Pi_{VS}(\mathcal{G})$ is equal to the cutsize of node partition $\Pi_{\mathcal{U}}(\mathcal{H})$ induced by $\Pi_{VS}(\mathcal{G})$ according to cutnet metric, whereas the separator size of $\Pi_{VS}(\mathcal{G})$ approximates the cutsize of node partition $\Pi_{\mathcal{U}}(\mathcal{H})$ induced by $\Pi_{VS}(\mathcal{G})$ according to the connectivity metric.*

Comparison of Figures 3.2(a) and 3.2(b) illustrates that the connectivities of separator vertices in $\Pi_{VS}$ are exactly equal to those of the cut nets of induced partial node partition $\Pi'_{\mathcal{U}}(\mathcal{H})$. Figure 3.2(b) shows a 3-way complete node partition $\Pi_{\mathcal{U}}(\mathcal{H})$ obtained by assigning the free nodes (shown with dashed lines) $u_{17}$ and $u_{18}$ to parts $\mathcal{U}_3$ and $\mathcal{U}_1$, respectively. This free node assignment does not increase the connectivities of the cut nets. However a different free node assignment might increase the connectivities of the cut nets. For example, assigning free node $u_{17}$ to part $\mathcal{U}_2$ instead of $\mathcal{U}_3$ will increase the connectivity of net $n_{15}$ by 1.

### 3.2.1. Recursive-bipartitioning-based partitioning.

In the recursive bipartitioning (RB) paradigm, a hypergraph is first partitioned into 2 parts. Then, each part of the bipartition is further bipartitioned recursively until the desired number of parts, $K$, is achieved. The following corollary forms the basis for the use of RB-based GPVS for RB-based HP according to the connectivity and the cut-net metrics.

COROLLARY 3. *Let $\Pi_{VS}(\mathcal{G}) = \{\mathcal{V}_1, \mathcal{V}_2; \mathcal{V}_S\}$ be a partition of $\mathcal{G}$ by a vertex separator $\mathcal{V}_S$, and let $\Pi_{\mathcal{U}}(\mathcal{H}) = \{\mathcal{U}_1, \mathcal{U}_2\}$ be a node partition of $\mathcal{H}$ that induces the net partition $\Pi_{\mathcal{N}}(\mathcal{H}) = \{\mathcal{N}_1 \equiv \mathcal{V}_1, \mathcal{N}_2 \equiv \mathcal{V}_2; \mathcal{N}_S \equiv \mathcal{V}_S\}$. The connectivity of a net $n_i$ in $\Pi_{\mathcal{U}}(\mathcal{H})$ is equal to the connectivity of the corresponding vertex $v_i$ in $\Pi_{VS}(\mathcal{G})$.*

**Separator-vertex removal:** In RB-based multiway HP, the cut-net metric is formulated by cut-net removal after each RB step. In this method, after each hypergraph bipartitioning step, each cut net is discarded from further RB steps. That is, a node bipartition $\Pi_{\mathcal{U}}(\mathcal{H}) = \{\mathcal{U}_1, \mathcal{U}_2\}$ of the current hypergraph $\mathcal{H}$, which induces the net bipartition $\Pi_{\mathcal{N}}(\mathcal{H}) = \{\mathcal{N}_1, \mathcal{N}_2; \mathcal{N}_S\}$, is decoded as generating two sub-hypergraphs $\mathcal{H}_1 = (\mathcal{U}_1, \mathcal{N}_1)$ and $\mathcal{H}_2 = (\mathcal{U}_2, \mathcal{N}_2)$ for further RB steps. Hence, the total cutsize of the resulting multiway partition of $\mathcal{H}$ according to the cut-net metric will be equal to the sum of the number of cut nets of the bipartition obtained at each RB step.

The cut-net metric can be formulated in the RB-GPVS-based multiway HP by separator-vertex removal so that each separator vertex is discarded from further RB steps. That is, at each RB step, a 2-way vertex separator $\Pi_{VS}(\mathcal{G}) = \{\mathcal{V}_1, \mathcal{V}_2; \mathcal{V}_S\}$ of $\mathcal{G}$ is decoded as generating two sub-graphs $\mathcal{G}_1 = (\mathcal{V}_1, \mathcal{E}_1)$ and $\mathcal{G}_2 = (\mathcal{V}_2, \mathcal{E}_2)$, where $\mathcal{E}_1$ and $\mathcal{E}_2$ denote the internal edges of the vertex parts $\mathcal{V}_1$ and $\mathcal{V}_2$, respectively. In other words, $\mathcal{G}_1$ and $\mathcal{G}_2$ are the sub-graphs of $\mathcal{G}$ induced by the vertex parts $\mathcal{V}_1$ and $\mathcal{V}_2$, respectively. $\mathcal{G}_1$ and $\mathcal{G}_2$ constructed in this way become the NIG representations of hypergraphs $\mathcal{H}_1$ and $\mathcal{H}_2$, respectively. Hence, the sum of the number of separator vertices of the 2-way GPVS obtained at each RB step will be equal to the total cutsize of the resulting multiway partition of $\mathcal{H}$ according to the cut-net metric.

**Separator-vertex splitting:** In RB-based multiway HP, the connectivity metric is formulated by adopting the cut-net splitting method after each RB step. In this method, each RB step, $\Pi_{\mathcal{U}}(\mathcal{H}) = \{\mathcal{U}_1, \mathcal{U}_2\}$ is decoded as generating two sub-hypergraphs $\mathcal{H}_1 = (\mathcal{U}_1, \mathcal{N}_1)$ and $\mathcal{H}_2 = (\mathcal{U}_2, \mathcal{N}_2)$ as in the cut-net removal method. Then, each cut net $n_s$ of $\Pi_{\mathcal{U}}(\mathcal{H})$ is split into two pin-wise disjoint nets $n_s^1$ and $n_s^2$ with $Pins(n_s^1) = Pins(n_s) \cap \mathcal{U}_1$ and $Pins(n_s^2) = Pins(n_s) \cap \mathcal{U}_2$, where $n_s^1$ and $n_s^2$ are added to the net lists of $\mathcal{H}_1$ and $\mathcal{H}_2$, respectively. In this way, the total cutsize of the resulting multiway partition according to the connectivity metric will be equal to the sum of the number of cut nets of the bipartition obtained at each RB step [9].

The connectivity metric can be formulated in the RB-GPVS-based multiway HP by separator-vertex splitting, which is not as easy as the separator-vertex removal method and it needs special attention. In a straightforward implementation of this method, a 2-way vertex separator $\Pi_{VS}(\mathcal{G}) = \{\mathcal{V}_1, \mathcal{V}_2; \mathcal{V}_S\}$ is decoded as generating two subgraphs $\mathcal{G}_1$ and $\mathcal{G}_2$ which are the sub-graphs of $\mathcal{G}$ induced by the vertex sets $\mathcal{V}_1 \cup \mathcal{V}_S$ and $\mathcal{V}_2 \cup \mathcal{V}_S$, respectively. That is, each separator vertex $v_s \in \mathcal{V}_S$ is split into two vertices $v_s^1$ and $v_s^2$ with $Adj(v_s^1) = Adj(v_s) \cap (\mathcal{V}_1 \cup \mathcal{V}_S)$ and $Adj(v_s^2) = Adj(v_s) \cap (\mathcal{V}_2 \cup \mathcal{V}_S)$. Then, the split vertices $v_s^1$ and $v_s^2$ are added to the subgraphs $(\mathcal{V}_1, \mathcal{E}_1)$ and $(\mathcal{V}_2, \mathcal{E}_2)$ to form $\mathcal{G}_1$ and $\mathcal{G}_2$, respectively.

This straightforward implementation of separator-vertex splitting method can be overcautious because of the unnecessary replication of separator edges in both subgraphs $\mathcal{G}_1$ and $\mathcal{G}_2$. Here an edge is said to be a separator edge if two vertices connected by the edge are both in the separator $\mathcal{V}_S$. Consider a separator edge $(v_{s_1}, v_{s_2}) \in \mathcal{E}$ in a given bipartition $\Pi_{VS}(\mathcal{G}) = \{\mathcal{V}_1, \mathcal{V}_2; \mathcal{V}_S\}$ of $\mathcal{G}$, where $\Pi_{\mathcal{U}}(\mathcal{H}) = \{\mathcal{U}_1, \mathcal{U}_2\}$ is a bipartition of $\mathcal{H}$ induced by $\Pi_{VS}(\mathcal{G})$ according to construction given in (3.5). If both $\mathcal{U}_1$ and $\mathcal{U}_2$ contain at least one node that induces the separator edge $(v_{s_1}, v_{s_2})$ of $\mathcal{G}$ then the replication of $(v_{s_1}, v_{s_2})$ in both subgraphs $\mathcal{G}_1$ and $\mathcal{G}_2$ is necessary. If, however, all hypergraph nodes that induce the edge $(v_{s_1}, v_{s_2})$ of $\mathcal{G}$ remain in only one part of $\Pi_{\mathcal{U}}(\mathcal{H})$ then the replication of $(v_{s_1}, v_{s_2})$ on the graph corresponding to the other part is unnecessary. For example, if all nodes connected by both nets $n_{s_1}$ and $n_{s_2}$ of $\mathcal{H}$ remain in $\mathcal{U}_1$ of $\Pi_{\mathcal{U}}(\mathcal{H})$ then the edge $(v_{s_1}, v_{s_2})$ should be replicated in only $\mathcal{G}_1$. $\mathcal{G}_1$ and $\mathcal{G}_2$ constructed in this way become the NIG representations of hypergraphs $\mathcal{H}_1$ and $\mathcal{H}_2$, respectively. Hence, the sum of the number of separator vertices of the 2-way GPVS obtained at each RB step will be equal to the total cutsize of the resulting multiway partition of $\mathcal{H}$ according to the connectivity metric.

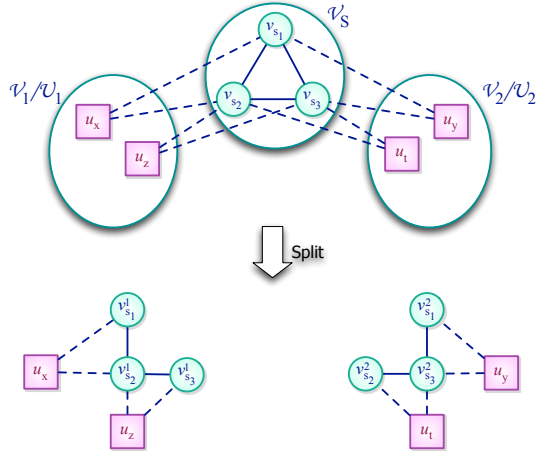

Fig. 3.3. *Separator-vertex splitting.*

Figure 3.3 illustrates three separator vertices $v_{s_1}$, $v_{s_2}$ and $v_{s_3}$ in a 2-way vertex separator and their splits into vertices $v_{s_1}^1, v_{s_2}^1, v_{s_3}^1$ and $v_{s_1}^2, v_{s_2}^2, v_{s_3}^2$. The three separator vertices $v_{s_1}$, $v_{s_2}$ and $v_{s_3}$ are connected with each other by three separator edges $(v_{s_1}, v_{s_2})$, $(v_{s_1}, v_{s_3})$ and $(v_{s_2}, v_{s_3})$ in order to show three distinct cases of separator edge replication in the accurate implementation. The figure also shows four hypergraph nodes $u_x, u_y, u_z$ and $u_t$ which induce the three separator edges, where $u_x, u_z$ are assigned to part $\mathcal{U}_1$ and $u_y, u_t$ are assigned to part $\mathcal{U}_2$. Since only $u_x$ induces the separator edge $(v_{s_1}, v_{s_2})$ and $u_x$ is assigned to $\mathcal{U}_1$, it is sufficient to replicate the separator edge $(v_{s_1}, v_{s_2})$ in only $\mathcal{V}_1$. Symmetrically, since only $u_y$ induces the separator edge $(v_{s_1}, v_{s_3})$ and $u_y$ is assigned to $\mathcal{U}_2$, it is sufficient to replicate the separator edge $(v_{s_1}, v_{s_3})$ in only $\mathcal{V}_2$. However, since $u_z$ and $u_t$ both induce the separator edge $(v_{s_2}, v_{s_3})$ and $u_z$ and $u_t$ are respectively assigned to $\mathcal{U}_1$ and $\mathcal{U}_2$, it necessary to replicate the separator edge $(v_{s_2}, v_{s_3})$ in both $\mathcal{V}_1$ and $\mathcal{V}_2$.

This accurate implementation of the separator-vertex splitting method depends on the availability of both $\mathcal{H}$ and its NIG representation $\mathcal{G}$ at the beginning of each RB step. Hence, after each RB step, the sub-hypergraphs $\mathcal{H}_1$ and $\mathcal{H}_2$ should be constructed as well as the subgraphs $\mathcal{G}_1$ and $\mathcal{G}_2$. We briefly summarize the details of the proposed implementation method performed at each RB step. A 2-way GPVS is performed on $\mathcal{G}$ to obtain a vertex separator $\Pi_{VS}(\mathcal{G})$. Then, a node bipartition $\Pi_{\mathcal{U}}(\mathcal{H})$ of $\mathcal{H}$ is constructed according to (3.5) by decoding the vertex separator $\Pi_{VS}(\mathcal{G})$ of $\mathcal{G}$. Then, the 2-way vertex separator $\Pi_{VS}(\mathcal{G})$ is used together with the node bipartition $\Pi_{\mathcal{U}}(\mathcal{H})$ to generate subgraphs $\mathcal{G}_1$ and $\mathcal{G}_2$ as described above. The sub-hypergraphs $\mathcal{H}_1$ and $\mathcal{H}_2$ are also constructed for the use in subsequent RB steps. An alternative implementation could be first generating sub-hypergraphs $\mathcal{H}_1$ and $\mathcal{H}_2$ from $\Pi_{\mathcal{U}}(\mathcal{H})$ and then constructing subgraphs $\mathcal{G}_1$ and $\mathcal{G}_2$ from $\mathcal{H}_1$ and $\mathcal{H}_2$, respectively, using NIG construction. However, this alternative implementation method is significantly inefficient compared to the proposed implementation, since construction of the NIG representation from a given hypergraph is computationally expensive.

**3.2.2. Balancing constraint.** Consider a node partition $\Pi_{\mathcal{U}}(\mathcal{H}) = \{\mathcal{U}_1, \mathcal{U}_2, \ldots, \mathcal{U}_K\}$ of $\mathcal{H}$ constructed from the vertex separator $\Pi_{VS}(\mathcal{G}) = \{\mathcal{V}_1, \mathcal{V}_2, \ldots, \mathcal{V}_K\}$ of NIG $\mathcal{G}$ according to (3.5). Since the vertices of $\mathcal{G}$ correspond to the nets of the given hypergraph $\mathcal{H}$, it is easy to enforce a balance criterion on the nets of $\mathcal{H}$ by setting $w(v_i) = w(n_i)$. For example, assuming unit net weights, the partitioning constraint of balancing on the vertex counts of parts of $\Pi_{VS}(\mathcal{G})$ infers balance among the internal net counts of node parts of $\Pi_{\mathcal{U}}(\mathcal{H})$.

However, balance on the nodes of $\mathcal{H}$ can not be directly enforced during the GPVS of $\mathcal{G}$, because the NIG model suffers from information loss on hypergraph nodes. Here, we propose a vertex-weighting model for estimating the cumulative weight of hypergraph nodes in each vertex part $\mathcal{V}_k$ of the vertex separator $\Pi_{VS}(\mathcal{G})$. In this model, the objective is to find appropriate weights for the vertices of $\mathcal{G}$ so that vertex-part weight $W(\mathcal{V}_k)$ computed according to (2.1) approximates the node-part weight $W(\mathcal{U}_k)$ computed according to (2.6).

The NIG model can also be viewed as a clique-node model since each node $u_h$ of the hypergraph induces an edge between each pair of vertices corresponding to the nets that connect $u_h$. So, the edges of $\mathcal{G}$ implicitly represent the nodes of $\mathcal{H}$. Each hypergraph node $u_h$ of degree $d_h$ induces $\binom{d_h}{2}$ clique edges among which the weight $w(u_h)$ is distributed evenly. That is, every clique edge induced by node $u_h$ can be considered as having a uniform weight of $w(u_h)/\binom{d_h}{2}$. Multiple edges between the same pair of vertices is collapsed into a single edge whose weight is equal to the sum of the weights of its constituent edges. Hence, the weight $w(e_{ij})$ of each edge $e_{ij}$ of $\mathcal{G}$ becomes,

$$w(e_{ij}) = \sum_{u_h \in Pins(n_i) \cap Pins(n_j)} \frac{w(u_h)}{\binom{d_h}{2}} \tag{3.6}$$

Then, the weight of each edge is uniformly distributed between the pair of vertices connected by that edge. That is, edge $e_{ij}$ contributes $w(e_{ij})/2$ to both $v_i$ and $v_j$. Hence, in the proposed model, the weight $w(v_i)$ of vertex $v_i$ becomes,

$$\begin{aligned} w(v_i) &= \frac{1}{2} \sum_{v_j \in Adj(v_i)} w(e_{ij}) \\ &= \sum_{u_h \in Pins(n_i)} \frac{w(u_h)}{d_h} \end{aligned} \tag{3.7}$$

Consider an internal hypergraph node $u_h$ of part $\mathcal{U}_k$ of $\Pi_{\mathcal{U}}(\mathcal{H})$. Since all graph vertices corresponding to the nets that connect $u_h$ are in part $\mathcal{V}_k$ of $\Pi_{VS}(\mathcal{G})$, $u_h$ will contribute $w(u_h)$ to $W(\mathcal{V}_k)$. Consider a boundary hypergraph node $u_h$ of part $\mathcal{U}_k$ with an external degree $\delta_h < d_h$, i.e., $u_h$ is connected by $\delta_h$ cut nets. Thus, $u_h$ will contribute by an amount of $(1-\delta_h/d_h)w(u_h)$ to $W(\mathcal{V}_k)$ instead of $w(u_h)$. So, vertex-part weight $W(\mathcal{V}_k)$ of $\mathcal{V}_k$ in $\Pi_{VS}(\mathcal{G})$ will be less than the actual node-part weight $W(\mathcal{U}_k)$ of $\mathcal{U}_k$ in $\Pi_{\mathcal{U}}(\mathcal{H})$. As the vertex-part weights of different parts of $\Pi_{VS}(\mathcal{G})$ will involve similar errors, the proposed method can be expected to produce a sufficiently good balance on the node-part weights of $\Pi_{\mathcal{U}}(\mathcal{H})$.

The free nodes can easily be exploited to improve the balance during the completion of partial node partition. For the cut-net metric in (2.8), we perform free-node-to-part assignment after obtaining $K$-way GPVS, since arbitrary assignments of free nodes do not disturb the cutsize by Corollary 2. However, for the connectivity metric in (2.9), free-node-to-part assignment needs special attention if it is performed after obtaining a $K$-way GPVS. According to Theorem 2, arbitrary assignments of free nodes may increase the connectivity of cut nets. So, for the connectivity cutsize metric, we perform free-node-to-part assignment after each RB step to improve the balance. Note that free-node-to-part assignment performed in this way does not increase the connectivity of cut nets in the RB-GPVS-based by Corollary 3. For both cutsize metrics, the best-fit-decreasing heuristic [50] used in solving the bin-packing problem is adapted to obtain a complete node partition/bipartition. Free nodes are assigned to parts in decreasing weight, where the best-fit criterion corresponds to assigning a free node to a part that currently has the minimum weight. Initial part weights are taken as the weights of the two parts in partial node bipartition.

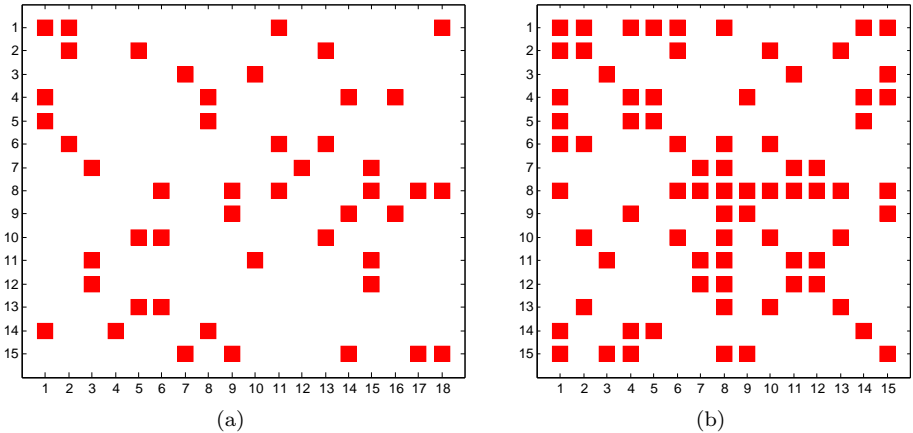

FIG. 3.4. *(a) A sample matrix $A$ whose row-net hypergraph representation $\mathcal{H}_A$ is equal to the sample hypergraph $\mathcal{H}$ given in Figure 3.1(a) and (b) the matrix $Z = AA^T$.*

**3.3. Matrix Theoretical View of the Relation Between HP and GPVS.** We will first briefly discuss the row-net and column-net models we proposed for representing rectangular as well as symmetric and nonsymmetric square matrices in our earlier work [8, 9, 45, 44]. These two models are duals: the row-net representation of a matrix is equal to the column-net representation of its transpose. Here, we only discuss the row-net model for permuting a matrix $A$ into a primal singly-bordered block-diagonal (SB) form, whereas the column-net model can be used for permuting $A$ into a dual SB form. In the row-net hypergraph model, an $M \times N$ matrix $A = (a_{ij})$ is represented as a hypergraph $\mathcal{H}_A = (\mathcal{U}, \mathcal{N})$ on $N$ nodes and $M$ nets with the number of pins equal to the number of nonzeros in matrix $A$. Node and net sets $\mathcal{U}$ and $\mathcal{N}$ correspond, respectively, to the columns and rows of $A$. There exist one net $n_i$ and one node $u_j$ for each row $i$ and column $j$, respectively. Net $n_i \subseteq \mathcal{U}$ contains the nodes corresponding to the columns that have a nonzero entry in row $i$, i.e., $u_j \in n_i$ if and only if $a_{ij} \neq 0$. That is, $Pins(n_i)$ represents the set of columns that have a nonzero in row $i$ of $A$, and in a dual manner $Nets(u_j)$ represents the set of rows that have a nonzero in column $j$ of $A$. Figure 3.4(a) shows an $15 \times 18$ matrix $A$ whose row-net hypergraph representation $\mathcal{H}_A$ is equal to the sample hypergraph $\mathcal{H}$ given in Figure 3.1(a).

Let $\mathcal{G}_{NIG}(\mathcal{H}_A) = (\mathcal{V}, \mathcal{E})$ denote the NIG model for the row-net hypergraph representation $\mathcal{H}_A = (\mathcal{U}, \mathcal{N})$ of matrix $A$. By definition of the NIG model, the vertices of $\mathcal{G}_{NIG}$ will represent the rows of $A$, and $e_{ij} \in \mathcal{E}$ if and only if $Pins(n_i) \cap Pins(n_j) \neq \emptyset$. Since $Pins(n_i)$ represents the set of columns that have a nonzero in row $i$ of $A$,

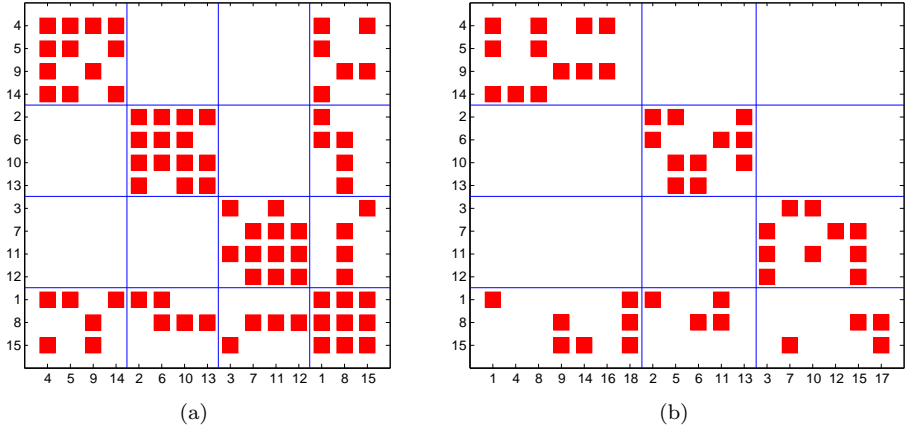

Fig. 3.5. *A 3-way DB from of the $AA^T$ matrix (a) SB form $A_{SB}$ of $A$ shown in Figure 3.4(a).*

$Pins(n_i) \cap Pins(n_j) \neq \emptyset$ corresponds to the condition that rows $i$ and $j$ of $A$ have a nonzero in at least one common column. Let $Z = (z_{ij})$ denote the $M \times M$ matrix $Z = AA^T$, an $\langle . \rangle$ denote inner-product operator. Since $z_{ij} = \langle r_i, r_j^T \rangle$, $z_{ij}$ will be nonzero if and only if $e_{ij} \in \mathcal{E}$. Hence, the sparsity pattern of symmetric matrix $Z$ will correspond to the adjacency matrix representation of $\mathcal{G}_{NIG}$. In other words, $\mathcal{G}_{NIG}$ will be equivalent to the standard graph representation of symmetric matrix $Z$, i.e., $\mathcal{G}_{NIG}(\mathcal{H}_A) \equiv \mathcal{G}_{AA^T}$. Note that although vertex $v_i$ of $\mathcal{G}_{NIG}$ represents only row $i$ of $A$, it represents both row $i$ and column $i$ of $AA^T$ in $\mathcal{G}_{AA^T}$.

Figure 3.4(b) shows the $15 \times 15$ matrix $Z = AA^T$. Note that the standard graph representation of $Z$ is equivalent to the NIG representation $\mathcal{G}_{NIG}(\mathcal{H}_A)$ of $\mathcal{H}_A$. As has long been used for nested dissection ordering for sparsity preserving factorizations, the problem of transforming a symmetric matrix into a DB form through symmetric row/column permutation can be modeled as a GPVS problem on its standard graph representation. So, Figure 3.5(a) shows a 3-way DB form of the $AA^T$ matrix induced by the 3-way GPVS $\Pi_{VS}(\mathcal{G})$ of $\mathcal{G}_{NIG}(\mathcal{H}_A)$ shown in Figure 3.4(b). Recall that the 3-way partition $\Pi_{\mathcal{U}}(\mathcal{H}_A)$ shown in Figure 3.1(b) is induced by $\Pi_{VS}(\mathcal{G})$. Hence, $\Pi_{VS}(\mathcal{G})$ induces the same SB form $A_{SB}$ of $A$ as shown in Figure 3.5(b).

**3.4. Multilevel implementation of GPVS-based HP formulation.** The state-of-the-art graph and hypergraph partitioning tools that adopt the multilevel framework, consist of three phases: *coarsening*, *initial partitioning*, and *uncoarsening*. In the first phase, a multilevel clustering is applied starting from the original graph/hypergraph by adopting various matching heuristics until the number of vertices in the coarsened graph/hypergraph reduces below a predetermined threshold value. Clustering corresponds to coalescing highly interacting vertices to supernodes. In the second phase, a partition is obtained on the coarsest graph/hypergraph using various heuristics including FM, which is an iterative refinement heuristic proposed for graph/hypergraph partitioning by Fiduccia and Mattheyses [24] as a faster implementation of the KL algorithm proposed by Kernighan and Lin [36]. In the third phase, the partition found in the second phase is successively projected back towards the original graph/hypergraph by refining the projected partitions on the intermediate level uncoarsened graphs/hypergraphs using various heuristics including FM.

One of the most important applications of GPVS is George's *nested–dissection* algorithm [25, 26], which has been widely used for reordering of the rows/columns of a symmetric, sparse, and positive definite matrix to reduce *fill* in the factor matrices. Here, GPVS is defined on the standard graph model of the given symmetric matrix. The basic idea in the nested dissection algorithm is to reorder symmetric matrix into a 2-way DB form so that no fill can occur in the off-diagonal blocks. The DB form of the given matrix is obtained through a symmetric row/column permutation induced by a 2-way GPVS. Then, both diagonal blocks are reordered by applying the dissection strategy recursively. The performance of the nested-dissection reordering algorithm depends on finding small vertex separators at each dissection step.

In this work, we adapted and modified the *onmetis* ordering code of *MeTiS* [31] for implementing our GPVS-based HP formulation. *onmetis* utilizes RB paradigm for obtaining multiway GPVS. Since $K$ is not known in advance for ordering applications, recursive bipartitioning operations continue until the weight of a part becomes sufficiently small. In our implementation, we terminate the recursive bipartitioning process whenever the number of parts become $K$.

The separator refinement scheme used in the uncoarsening phase of *onmetis* considers vertex moves from vertex separator $\Pi_{VS}(\mathcal{G})$ to both $\mathcal{V}_1$ and $\mathcal{V}_2$ in $\Pi_{VS} = \{\mathcal{V}_1, \mathcal{V}_2; \mathcal{V}_S\}$. During these moves, *onmetis* uses the following feasibility constraint, which incorporates the size of the separator in balancing, i.e.,

$$\max\{W(\mathcal{V}_1), W(\mathcal{V}_2)\} \leq (1+\epsilon)\frac{W(\mathcal{V}_1)+W(\mathcal{V}_2)+W(\mathcal{V}_S)}{2} = W_{max} \qquad (3.8)$$

However, this may become a loose balancing constraint compared to (2.2) for relatively large separator sizes which is typical during refinements of coarser graphs. This loose balancing constraint is not an important concern in *onmetis*, because it is targeted for fill-reducing sparse matrix ordering which is not very sensitive to the imbalance between part sizes. Nevertheless, this scheme degrades the load balancing quality of our GPVS-based HP implementation, where load balancing is more important in the applications for which HP is utilized. We modified *onmetis* by computing the maximum part weight constraint as

$$W_{max} = (1+\epsilon)(W(\mathcal{V}_1) + W(\mathcal{V}_2))/2 \qquad (3.9)$$

at the beginning of each FM pass, whereas *onmetis* computes $W_{max}$ according to (3.8) once for all FM passes, in a level. Furthermore, *onmetis* maintains only one value for each vertex which denotes both the weight and the cost of the vertex. We added a second field for each vertex to hold the weight and the cost of the vertex separately. The weights and the costs of vertices are accumulated independently during vertex coalescings performed by matchings at the coarsening phases. Recall that weight values are used for maintaining the load balancing criteria, whereas cost values are used for computing the size of the separator. That is, FM gains of the separator vertices are computed using the cost values of those vertices.

The GPVS-based HP implementation obtained by adapting *onmetis* as described in this subsection will be referred to as *onmetisHP*.

**4. Experimental Results.** We test the performance of our GPVS-based HP formulation by partitioning matrices from the linear-programming (LP) and the positive definite (PD) matrix collections of the University of Florida matrix collection [18]. Matrices in the latter collection are square and symmetric, whereas the matrices in the former collection are rectangular. The row-net hypergraph models [9, 13] of the test matrices constitute our test set. In these hypergraphs, nets are associated with unit cost. To show the validity of our GPVS-based HP formulation, test hypergraphs are partitioned by both *PaToH* and *onmetisHP* and default parameters are utilized in both tools. In general, the maximum imbalance ratio $\epsilon$ was set to be 10%.

We excluded small matrices that have less than 1000 rows or 1000 columns. In the LP matrix collection, there were 190 large matrices out of 342 matrices. Out of these 190 large matrices, 5 duplicates, 1 extremely large matrix and 5 matrices, for which NIG representations are extremely large were excluded. We also excluded 26 outlier matrices which yield large separators[1] to avoid skewing the results. Thus, 153 test hypergraphs are used from the LP matrix collection. In the PD matrix collection, there were 170 such large matrices out of 223 matrices. Out of these 170 large matrices, 2 duplicates, 2 matrices, for which NIG representations are extremely large and 7 matrices with large separators were excluded. Thus, 159 test hypergraphs are used from the PD matrix collection. We experimented with $K$-way partitioning of test hypergraphs for $K = 2, 4, 8, 16, 32, 64$, and 128. For a specific $K$ value, $K$-way partitioning of a test hypergraph constitutes a partitioning instance. For the LP collection, instances in which $min\{|\mathcal{U}|, |\mathcal{N}|\} < 50K$ are discarded as the parts would become too small. So, 153, 153, 153, 153, 135, 100, and 65 hypergraphs are partitioned for $K = 2, 4, 8, 16, 32, 64$, and 128, respectively, for the LP collection. Similarly for the PD collection, instances in which $|\mathcal{U}| < 50K$ are discarded. So, 159, 159, 159, 159, 145, 131, and 109 hypergraphs are partitioned for $K = 2, 4, 8, 16, 32, 64$, and 128, respectively for the PD collection. In this section, we summarize our findings in these experiments. Please refer to [35] for detailed experimental results for each partitioning instance.

The hypergraphs obtained from the LP matrix collection are used for permuting the matrices into singly-bordered (SB) block-angular-form for coarse-grain parallelization of linear-programming applications [3]. Here, minimizing the cutsize according to the cut-net metric (2.4) corresponds to minimizing the size of the row border in the induced SB form. In these applications, nets are either associated with unit weights or weights that are equal to the nonzeros in the respective rows. In the former case, net balancing corresponds to balancing the row counts of the diagonal blocks, whereas in the latter case, net balancing corresponds to balancing the nonzero counts of the diagonal blocks. Experimental comparisons are provided only for the former case, because *PaToH* does not support different cost and weight associations to nets.

The hypergraphs obtained from the PD matrix collection are used for minimizing communication overhead in column-parallel matrix-vector multiply algorithm in iterative solvers. Here, minimizing the cutsize according to the connectivity metric (2.5) corresponds to minimizing the total communication volume when the point-to-point inter-processor communication scheme is used [9]. Minimizing the cutsize according to the cut-net metric (2.4) corresponds to minimizing the total communication volume when the collective communication scheme is used [13]. In these applications, nodes associated with weights that are equal to the nonzeros in the respective columns. So, balancing part weights corresponds to computational load balancing.

In the following tables, the performance figures are computed and displayed as follows. Since both *PaToH* and *onmetisHP* tools involve randomized heuristics, 10 different partitions are obtained for each partitioning instance and the geometric average of the 10 resultant partitions are computed as the representative results for both HP tools on the particular partitioning instance. For each partitioning instance, the cutsize value is normalized with respect to the total number of nets in the respective hypergraph. Recall that all test hypergraphs have unit-cost nets. So, for the cut-net metric, these normalized cutsize values show the fraction of the cut nets. For the connectivity metric, these normalized cutsize values show the average net connectivity. For each partitioning instance, the running time of $PaToH$ is normalized with respect to that of $onmetisHP$, thus showing the speedup obtained by $onmetisHP$ for that partitioning instance. These normalized cutsize values and speedup values as well as percent load imbalance values are summarized in the tables by taking the geometric averages for each $K$ value.

---

[1]Here, a separator is said to be large if it includes more than 33% of all nets.

TABLE 4.1
*Performance averages on the LP matrix collection for cut-net metric with net balancing.*

| | $PaToH$ | | $onmetisHP$ | | |
|---|---|---|---|---|---|
| $K$ | cutsize | %LI | cutsize | %LI | speedup |
| 2 | 0.02 | 1.2% | 0.03 | 0.3% | 2.04 |
| 4 | 0.02 | 1.9% | 0.05 | 2.6% | 2.45 |
| 8 | 0.07 | 3.1% | 0.09 | 6.9% | 2.64 |
| 16 | 0.09 | 5.2% | 0.14 | 13.0% | 2.78 |
| 32 | 0.13 | 8.8% | 0.18 | 23.1% | 2.83 |
| 64 | 0.15 | 11.5% | 0.21 | 27.8% | 2.83 |
| 128 | 0.16 | 13.5% | 0.21 | 31.3% | 2.76 |

Table 4.1 displays overall performance averages of $onmetisHP$ compared to those of $PaToH$ for the cut-net metric in (see (2.8)) with net balancing on the LP matrix collection. As seen in Table 4.1, $onmetisHP$ obtains hypergraph partitions of comparable cutsize quality with those of $PaToH$. However, load balancing quality of partitions produced by $onmetisHP$ is worse than those of $PaToH$, especially with increasing $K$. As seen in the table, $onmetisHP$ runs significantly faster than $PaToH$ for each $K$. For example, $onmetisHP$ runs 2.83 times faster than $PaToH$ for 32-way partitionings on the average.

TABLE 4.2
*Performance averages on the PD matrix collection for cut-net metric with node balancing.*

| | $PaToH$ | | $onmetisHP$ | | | | |
|---|---|---|---|---|---|---|---|
| $K$ | cutsize | %LI | cutsize | $exp\%LI_p$ | $act\%LI_p$ | $act\%LI_c$ | speedup |
| 2 | 0.01 | 0.1% | 0.01 | 0.2% | 0.2% | 0.1% | 1.40 |
| 4 | 0.03 | 0.3% | 0.03 | 0.9% | 1.5% | 1.1% | 1.75 |
| 8 | 0.05 | 0.4% | 0.05 | 2.8% | 3.7% | 2.7% | 1.96 |
| 16 | 0.08 | 0.6% | 0.08 | 6.7% | 7.4% | 5.4% | 1.98 |
| 32 | 0.12 | 0.9% | 0.12 | 13.4% | 12.8% | 9.2% | 2.17 |
| 64 | 0.17 | 1.2% | 0.16 | 22.1% | 19.8% | 13.5% | 2.27 |
| 128 | 0.25 | 1.6% | 0.24 | 32.5% | 28.8% | 17.9% | 2.25 |

Table 4.2 displays overall performance averages of $onmetisHP$ compared to those of $PaToH$ for the cut-net metric with node balancing on the PD matrix collection. In the table, $exp\%LI_p$ and $act\%LI_p$ respectively denote the expected and actual percent load imbalance values for the partial node partitions of the hypergraphs induced by $K$-way GPVS. $act\%LI_c$ denotes the actual load imbalance values for the complete node partitions obtained after free-node-to-part assignment. The small discrepancies between the $exp\%LI_p$ and $act\%LI_p$ values show the validity of the approximate weighting scheme proposed in Section 3.2 for the vertices of the NIG. As seen in the table, for each $K$, the $act\%LI_c$ value is considerably smaller than the $act\%LI_p$ value. This experimental finding confirms the effectiveness of the free-node-to-part assignment scheme mentioned in Section 3.2. As seen in Table 4.2, $onmetisHP$ obtains hypergraph partitions of comparable cutsize quality with those of $PaToH$. However, load balancing quality of partitions produced by $onmetisHP$ is considerably worse than those of $PaToH$. As seen in the table, $onmetisHP$ runs considerably faster than $PaToH$ for each $K$.

TABLE 4.3
*Comparison of accurate and overcautious separator-vertex splitting implementations with averages on the PD matrix collection for connectivity metric with node balancing.*

| | overcautous / accurate | | |
|---|---|---|---|
| $K$ | cutsize | %LI | speedup |
| 2 | 1.00 | 0.63 | 1.29 |
| 4 | 1.02 | 0.79 | 1.50 |
| 8 | 1.10 | 0.79 | 1.61 |
| 16 | 1.29 | 0.70 | 1.63 |
| 32 | 1.56 | 0.64 | 1.61 |
| 64 | 1.84 | 0.69 | 1.60 |
| 128 | 2.09 | 0.60 | 1.54 |

Table 4.3 is constructed based on the PD matrix collection to show the validity of the accurate vertex splitting formulation proposed in Section 3.2.1 for the connectivity cutsize metric (see ( 2.9)). In this table, speedup, cutsize and load imbalance values of $onmetisHP$ that uses the straightforward (overcautious) separator-vertex splitting implementation are normalized with respect to those of $onmetisHP$ that uses the accurate implementation. In the straightforward implementation, free-node-to-part assignment is performed after obtaining a $K$-way GPVS, since hypergraphs are not carried through the RB process. Free nodes are assigned to parts in decreasing weight, where the best-fit criterion corresponds to assigning a free node to a part which increases connectivity cutsize by the smallest amount with ties are broken in favor of the part with minimum weight. As seen in the table, the overcautious implementation leads to slightly better load balance than accurate implementation, because overcautious implementation performs free-node-to-part assignment on the $K$-way partial node partition induced by the $K$-way GPVS. As also seen in the table, the overcautious implementation, as expected, leads to slightly better speedup than the accurate implementation. However, the accurate implementation leads to significantly less cutsize values.

TABLE 4.4
*Performance averages on the PD matrix collection for connectivity metric with node balancing.*

| | $PaToH$ | | $onmetisHP$ | | |
|---|---|---|---|---|---|
| $K$ | cutsize | %LI | cutsize | %LI | speedup |
| 2 | 1.03 | 0.1% | 1.03 | 0.2% | 1.29 |
| 4 | 1.08 | 0.3% | 1.08 | 0.8% | 1.50 |
| 8 | 1.15 | 0.5% | 1.15 | 1.7% | 1.61 |
| 16 | 1.26 | 0.7% | 1.25 | 4.1% | 1.63 |
| 32 | 1.37 | 1.0% | 1.36 | 7.9% | 1.61 |
| 64 | 1.49 | 1.5% | 1.47 | 11.8% | 1.60 |
| 128 | 1.63 | 1.9% | 1.60 | 16.5% | 1.54 |

Table 4.4 displays overall performance averages of $onmetisHP$ compared to those of $PaToH$ for the connectivity metric with node balancing on the PD matrix collection. In contrast to Table 4.2, load imbalance values are not displayed for partial node partitions in Table 4.4, because free-node-to-part assignments are performed after each 2-way GPVS operation for the sake of accurate implementation of the separator-vertex splitting method as mentioned in Section 3.2. So, $\%LI$ values displayed in Table 4.4 show the actual percent imbalance values for the $K$-way node partitions obtained. As seen in Table 4.4, similar to results of Table 4.2, $onmetisHP$ obtains hypergraph partitions of comparable cutsize quality with those of $PaToH$, whereas load balancing quality of partitions produced by $onmetisHP$ is considerably worse than those of $PaToH$. As seen in Table 4, $onmetisHP$ still runs considerably faster than $PaToH$ for each $K$ for the connectivity metric. However, the speedup values in Table 4.4 , are considerable smaller compared to those displayed in Table 2, which is due to the fact that $onmetisHP$ carries hypergraphs during the RB process for the sake of accurate implementation of the separator-vertex splitting method as mentioned in Section 3.2.

A common property of Tables 4.1, 4.2, and 4.4, is the increasing speedup of $onmetisHP$ compared to $PaToH$ with increasing $K$ values. This experimental finding stems from the fact that the initial NIG construction overhead amortizes with increasing $K$. Another common property of Tables 4.1, 4.2, and 4.4, is that $onmetisHP$ runs significantly faster than $PaToH$, while producing partitions of comparable cutsize quality with, however, worse load balancing quality. These experimental findings justify our GPVS-based hypergraph partitioning formulation for effective parallelization of applications in which computational balance definition is not very precise and preprocessing overhead due to partitioning overhead is important.

**5. Conclusions.** We have presented how the hypergraph partitioning problem can be efficiently and effectively solved through finding vertex separators on the net intersection graph representation of a hypergraph. Our empirical study on a wide set of test matrices showed that runtimes can be as much as 4.17 times faster, with the cutsize quality is preserved on average (and improved in many cases), while balance was achieved for small number of parts and remained acceptable for large number of parts. Moreover, we proposed techniques that can trade off cutsize and runtime against balance, showing that balance can be achieved even for high number of parts. Overall results prove that, the proposed hypergraph partitioning through vertex separators on graphs is ideal for applications where balance is not well-defined, which is the main motivation for our work, and competitive for application where balance is important.

We believe that the success of the proposed methods point to several future research directions. First, better vertex weighting schemes to approximate the node balance is an area that can make a significant impact. We believe exploiting domain specific information or devising techniques that can apply to certain classes of graphs, as opposed to constructing generic approximations that can work for all graphs, is a promising avenue to explore. Secondly, the algorithms we have used in this paper, were only slightly adjusted for the particular problem we were solving. There is a lot of room for improvements in algorithms for finding vertex separators with balanced hypergraph partitions, and we believe these algorithms can be designed and implemented on the existing partitioning graph partitioning frameworks, which means strong algorithmic ideas can be translated into effective software tools with relatively little effort. Finally, this paper is only an example of the growing importance of graph partitioning and the need for more flexible models for graph partitioning. Graph partitioning now is an internal step for divide-and-conquer based methods, whose popularity will only increase with the growing problems sizes. As such, re-quirements for graph partitioning will keep growing and broadening. While, the state of the art for graph partitioning has drastically improved from the days of merely minimizing the number of cut edges, while keeping the number of vertices balanced between the *two* parts, we believe there is still a lot of room for growth for more general models for graph partitioning.